\documentclass[aps,prl,10pt,superscriptaddress, showpacs, preprintnumbers, nofootinbib, 
twocolumn]{revtex4-1}
\pdfoutput=1
\usepackage{hyperref}
\usepackage{graphicx}
\usepackage{enumerate}
\usepackage{amsfonts,amsmath,amssymb,bm,bbm}
\usepackage{color}
\usepackage{cancel}
\usepackage{epsfig, subfigure}
\usepackage{setspace}
\usepackage{booktabs, tabularx}
\usepackage{slashed} 
\usepackage{units}

\hypersetup{
    pdfnewwindow=true,      
    colorlinks=true,       
    linkcolor=blue,          
    citecolor=blue,        
    filecolor=blue,      
    urlcolor=blue        
}

\newcommand{\be}{\begin{equation}}  
\newcommand{\ee}{\end{equation}}
\newcommand{\bea}{\begin{eqnarray}}  
\newcommand{\eea}{\end{eqnarray}}

\begin{document}

\title{WIMP Dark Matter and Neutrino Mass from Peccei-Quinn Symmetry}

\author{Basudeb Dasgupta}
\email{bdasgupta@ictp.it}
\affiliation{International Centre for Theoretical Physics, Strada Costiera 11, 
34014 Trieste, Italy.}

\author{Ernest Ma}
\email{ernest.ma@ucr.edu}
\affiliation{Department of Physics and Astronomy, University of California, 
Riverside, California 92521, USA.}

\author{Koji Tsumura}
\email{ko2@eken.phys.nagoya-u.ac.jp}
\affiliation{Department of Physics, Graduate School of Science, Nagoya 
University, Nagoya 464-8602, Japan.}

\date{December 16, 2013}
\begin{abstract}
The Peccei-Quinn anomalous global $U(1)_{PQ}$ symmetry is important for 
solving the strong CP problem with a cosmologically relevant axion. 
We add to this the simple (but hitherto unexplored) observation that it 
also has a residual $Z_2$ symmetry which may be responsible for 
a second component of dark matter, i.e. an absolutely stable 
weakly interacting singlet scalar.  This new insight provides a 
theoretical justification for this well-studied simplest of all 
possible dark-matter models.  It also connects with the well-studied 
notion of generating radiative neutrino mass through dark matter. Two 
such specific realizations are proposed.  In our general scenario, 
dark-matter detection is guaranteed at existing direct-detection 
experiments or axion searches.  Observable signals at the Large Hadron 
Collider are discussed.
\end{abstract}

\preprint{Preprint No: UCRHEP-T534}

\maketitle

{\bf{\emph{Introduction.---}}}
The standard model (SM) of particle interactions is missing at least three 
important pieces: (1) a natural explanation of the absence or suppression 
of strong CP violation, (2) the existence of dark matter (DM), and (3) the 
presence of neutrino mass.  The best motivated solution to the strong CP problem 
is the well-known Peccei-Quinn anomalous global $U(1)_{PQ}$ 
symmetry~\cite{Peccei:1977hh} which predicts a very light pseudoscalar particle -- 
the axion~\cite{Weinberg:1977ma,Wilczek:1977pj}, which may very well also be the 
DM.  An elegant way to get small neutrino masses is the seesaw mechanism
\,(see Ref.\,\cite{Mohapatra:2006gs} for a review), 
which postulates heavy neutral singlet fermions coupling to 
the observed neutrinos, elevating their masses from zero to small nonzero values.

Recognizing that the $U(1)_{PQ}$ breaking scale and the seesaw scale are 
both very high, say $10^{10}$\,GeV, it was proposed some years 
ago~\cite{Ma:2001ac} that they may be related in the context of supersymmetry.  
In that scenario, the lightest neutralino (which may be the axino) is also 
a DM candidate.  DM has thus two components.  The very 
light axion is not absolutely stable, but has a lifetime much longer than 
that of the Universe.  The heavy neutralino is absolutely stable because 
of the usual $R$ parity from supersymmetry, and it behaves as a 
Weakly Interacting Massive Particle (WIMP) in the usual cold DM scenario.

In this paper, we show that $U(1)_{PQ}$ can address all the three 
deficiencies of the SM without invoking supersymmetry. The 
$U(1)_{PQ}$ symmetry not only cures the strong CP problem (1), but it is also
 the origin of a previously unidentified residual $Z_2$ symmetry that may be responsible for 
a heavy second component of DM (2) which is absolutely stable, 
as well as radiative neutrino mass (3).

There are three generic realistic implementations of $U(1)_{PQ}$, differing mainly in
the choice of colored fermions charged under $U(1)_{PQ}$. In the 
KSVZ model~\cite{Kim:1979if,Shifman:1979if}, new heavy electroweak 
singlet quarks transforming under $U(1)_{PQ}$ are added. In the DFSZ 
model~\cite{Dine:1981rt,Zhitnitsky:1980tq} the regular quarks are chosen to transform 
under $U(1)_{PQ}$, but additional Higgs fields are added. In the gluino 
axion model~\cite{Demir:2000ng}, supersymmetry is assumed and 
$U(1)_{PQ}$ is identified with $U(1)_R$ so that gluinos are the only colored 
fermions transforming under $U(1)_{PQ}$. All these three realizations satisfactorily 
explain the smallness of the strong CP violation~\cite{Kim:2008hd, Kawasaki:2013aaa}. 
We now proceed to explain DM and neutrino masses.

We start with the simple (but hitherto unexplored) observation that in all these axion 
models, the spontaneous breaking of $U(1)_{PQ}$ actually also leaves a discrete $Z_2$ 
symmetry which is exactly conserved (see \cite{Krauss:1988zc, Weinberg:2013kea, Lindner:2011it} for some related ideas).  In the DFSZ model, it is $(-1)^{3B}$, where $B$ is baryon number.  In the gluino axion model, it is $R$ parity. In the KSVZ model 
it is a new symmetry distinguishing the heavy singlet quarks and any additional 
particles charged under $U(1)_{PQ}$ from all other particles. Hence the 
lightest new heavy neutral particle, odd under the $Z_2$ 
symmetry, will be absolutely stable and a potential WIMP candidate for DM. Similarly, 
neutrino mass terms may be forbidden at tree level by this same $Z_2$ symmetry 
and arise only radiatively~\cite{Ma:2006km}.  This new residual Peccei-Quinn 
$Z_2$ symmetry is thus tailor-made for having an absolutely stable DM 
component (in addition to the axion) and realizing the notion that neutrino 
mass is induced radiatively by DM.  Note that we do not have to introduce an 
extra symmetry by hand; it is already built into the axion model.

In the following, we will first present the simplest implementation of the above mechanism 
in the KSVZ model, to provide a stable heavy DM candidate and discuss its 
phenomenology.  Then we elaborate on two specific models of radiative neutrino 
mass derived from the above, together with the associated new particles and their 
collider phenomenology.

{\bf{\emph{WIMPs in Axion Models.---}}}
Consider the KSVZ model, using a heavy singlet quark $Q$ of charge $-1/3$ for 
the color anomaly which generates 
the axion.  Note that the domain wall number is one in this case, so 
the model is cosmologically safe~\cite{Sikivie:1982qv}.  We add a neutral 
complex singlet scalar $\chi$, which transforms under $U(1)_{PQ}$, to 
provide a heavy DM candidate. The axion 
is contained in the scalar field $\zeta$ which couples to $\bar{Q}Q$, 
and $\chi \chi$. Consider the Lagrangian relevant for $Q_{L,R}$, $\zeta$, 
and $\chi$, 
\begin{align}
{\cal L} =& \mu^2_\zeta |\zeta|^2 + {1 \over 2} \lambda_\zeta |\zeta|^4 + 
\mu_\chi^2 |\chi|^2 + {1 \over 2} \lambda_\chi |\chi|^4 
+ \lambda' |\zeta|^2 |\chi|^2 \nonumber \\
&\! + \bigl\{ f_Q \zeta \bar{Q}_L Q_R + f_d \chi \bar{Q}_L d_R 
+ \epsilon_\chi \zeta^* \chi \chi + \text{H.c.} \bigr\},
\label{Eq:AxionLag}
\end{align}
where $\chi = (\chi_1^{} + i\chi_2^{})/\sqrt{2}$. 
Let $\zeta = e^{ia/F_a}(F_a + \sigma)/\sqrt{2}$, where $a$ is the axion 
and $F_a= \sqrt{-2 \mu_\zeta^2/ \lambda_\zeta}$, the vacuum expectation value (VEV) 
that also acts as the axion decay constant.
 
In general, in axion models $U(1)_{PQ}$ is broken by the VEV of a scalar that couples to some $\bar{Q}_L Q_R$ (e.g., the $1^{\rm st}$ term on $2^{\rm nd}$ line in Eq.\,\ref{Eq:AxionLag}). After $U(1)_{PQ}$ symmetry breaking, one finds that $(\sigma,\,a)\rightarrow+(\sigma,\,a)$ and $Q_{L,R}\rightarrow\pm\,Q_{L,R}$ is a residual symmetry of the Lagrangian. Thus, ${\cal L}$ has a $Z_2$ symmetry under which $\sigma$ and $a$ must be even, whereas the particle $Q$ is odd, as also in~\cite{Krauss:1988zc}. If the fermion $Q$ were a known fermion, e.g., a regular quark for the DFSZ model or a gluino for the gluino axion model, the $Z_2$ would be identified with $(-1)^{3B}$ or $R$ parity, respectively. As $Q$ is a new fermion, this $Z_2$ is a new symmetry, say ``$Q$-parity''. The complex scalar $\chi$ is also forced to be odd under $Q$-parity (by the $2^{\rm nd}$ term on $2^{\rm nd}$ line in Eq.\,\ref{Eq:AxionLag}), thus stabilizing it (unless $d$ is charged, which would take us back to the DFSZ model). $Q$-parity must be \emph{exactly} preserved, otherwise the axion solution to the strong-CP problem is spoiled.

Assuming $\epsilon_\chi$ to be real for simplicity, the
mass eigenvalues of $\chi$ are $m_{1,2}^2 = \mu_\chi^2 + (1/2) \lambda' F_a^2 
\pm \epsilon_\chi F_a \sqrt{2}$.  Without loss of generality, 
we choose $\epsilon_\chi < 0$ and find that $m_1^{} < m_2^{}$, so that
then $\chi_1^{}$ could be DM. Since $F_a > 4 \times 10^8$ GeV from 
supernova SN1987A data~\cite{Raffelt:1987yt}, fine tuning is unavoidable for 
$m_{1,2} \sim$ TeV.  However, this problem plagues all (nonsupersymmetric) axion 
models because the electroweak Higgs doublet also has a large 
quantum correction.  On the other hand, there is a justification for 
$\epsilon_\chi$ to be small, 
from the fact that the limit $\epsilon_\chi = 0$ corresponds to an extra 
$U(1)$ symmetry, i.e., $\chi, Q_L, Q_R \sim 1$ independent of $U(1)_{PQ}$. 
The heavy KSVZ quark $Q$ with $m_Q^{} = f_Q^{} F_a/\sqrt{2}$ may also be 
observable if $m_Q^{} \sim$ TeV, i.e. $f_Q^{} \sim 10^{-6}$ for $F_a \sim 10^9$ GeV.

The are, therefore, two DM candidates in this model -- a light ultracold axion $a$, 
and a heavy cold WIMP-like $\chi_1^{}$.  The total cosmological DM density is the sum of 
their densities, i.e., $\Omega_{\rm DM}=\Omega_{a} + \Omega_{\chi_1^{}}$. 
The axion is massless until color chiral symmetry breaking, and it gets a mass 
$m_a\approx6\mu{\rm eV}(10^{12}{\rm\,GeV}/F_a)$~\cite{Preskill:1982cy, Abbott:1982af, Dine:1982ah}. 
For reheating temperatures lower than $F_a$, the only process relevant for 
axion production is coherent oscillation due to vacuum misalignment~\cite{Sikivie:2006ni}. 
The axion density is given by~\cite{Bae:2008ue} 
\begin{equation}
{\Omega_{a}h^2}\approx 0.18\,\theta_a^2\left(\frac{F_a}{10^{12}
{\rm\,GeV}}\right)^{1.19}\,,
\end{equation} 
where $\theta_a$ is the initial axion misalignment angle.
 
The WIMP DM candidate $\chi_1^{}$ has two main interactions with 
SM particles -- with down-type quarks through 
$f_d\bar{Q}_Ld_R^{}\chi$, and with the SM Higgs boson $h$ through the 
$\lambda_{\chi h} \chi^2 (\Phi^\dagger \Phi) \rightarrow (1/4) \lambda_{\chi h}\chi_1^2 (v_{\rm SM}^{} +h)^2$ term.
The annihilation cross section to down-type quark pairs is 
$\langle \sigma v\rangle\approx3f_d^4 m_d^2/(16 \pi (m_Q^2+m_1^2)^2)$, which for 
$m_{Q}$ and $m_1$\,$\sim$\,TeV turns out to be too small by a few orders of magnitude 
to yield the correct relic density.  The true $\chi_1^{}$ abundance is then  
set by the chemical freeze-out of its annihilation processes through the 
Higgs coupling.  However there is also the nonthermal production of $\chi$ 
from the decay of the radial field $\sigma$ which may be significant. 
This potential problem is absent in our model because the $\bar{Q}_Ld_R^{}\chi$ 
interaction, already built into the model, helps to keep $Q$, $\chi$, and $d$ in thermal 
equilibrium until late times, so that any nonthermal population 
of $\chi_1^{}$ is quickly rethermalized. 
Our scenario is then identical to that of the scalar singlet DM 
model~\cite{Silveira:1985rk, McDonald:1993ex, Burgess:2000yq}, and our results provide a 
theoretical justification of this well-studied simplest of all possible dark-matter models. 
The phenomenology of this model was recently updated  
in Ref.\,\cite{Cline:2013gha}, and we can directly use the results and 
constraints therein.

The relic abundance of $\chi_1^{}$ is determined by its coupling to the Higgs. 
For a heavy DM, $m_1>$ few\,$\times100\,{\rm GeV}$, the cross section simply 
goes as $\lambda_{\chi h}^2/m_1^2$ and an annihilation cross section of 
$\langle \sigma v\rangle\approx {\rm few}\times 10^{-26}\,{\rm cm}^3{\rm s}^{-1}$\,\cite{Steigman:2012nb} 
may be achieved quite easily. The relic density of DM in this case is approximately 
fit by~\cite{Cline:2013gha}
\begin{equation}
\frac{\Omega_{\chi_1^{}}}{\Omega_{\rm DM}}\approx 4\times10^{-7}\frac{(m_1/{\rm GeV})^2}
{\lambda_{\chi h}^2}\,.
\end{equation}

\begin{figure}[!t]
\centering
\includegraphics[angle=0.0, width=0.37\textwidth]{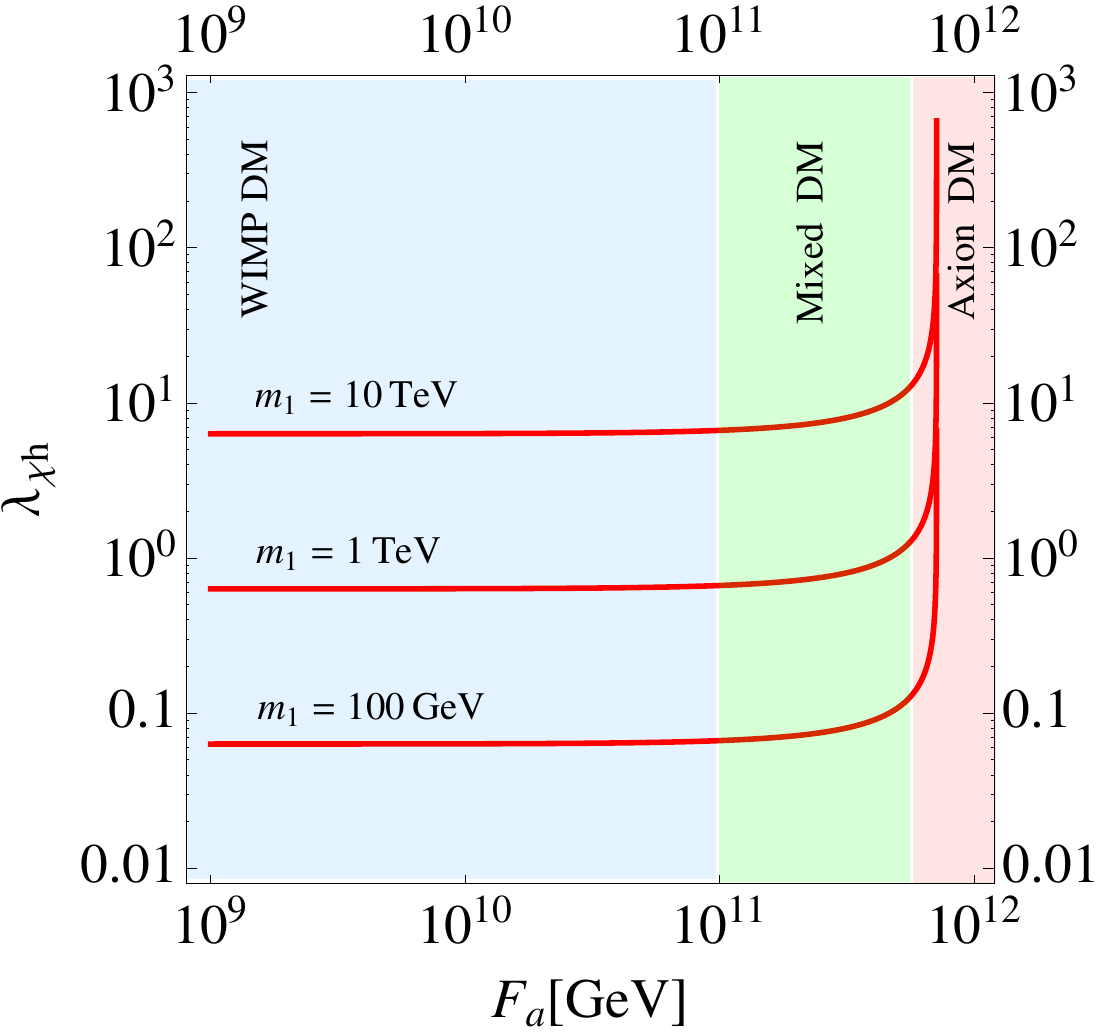}
\caption{Correlated values of WIMP-Higgs coupling $\lambda_{\chi h}$ and axion decay constant 
$F_a$ for various DM masses $m_1$, so that the total DM density in axions and $\chi_1$ is 
the observed value $\Omega_{\rm DM} h^2=0.12$. For concreteness, $\theta_a=1$ is assumed.}
\label{fig:mixeddm}
\end{figure}

Our scenario is related to the mixed axion-neutralino models 
reviewed in Ref.~\cite{Bae:2013pxa} (see references therein for details). Interestingly, 
although $\sigma$ imitates the role of the saxion, we have an inbuilt mechanism to 
keep $\sigma$ decay products in equilibrium, first by equilibrating them 
with the heavy quarks and then through color interactions with the SM quarks.
This allows us to consider the simplified DM production discussed 
above.  
However, more careful treatment may be needed in some cases, e.g., if the $\sigma$ 
decays to axions become important~\cite{Bae:2013qr}. Then one has to solve 
several coupled Boltzmann equations to study the model in detail.
It should be noted, however, that our insight into the hidden $Z_2$ symmetry of axion 
models provides a general mechanism for mixed axion-WIMP DM, independent of 
supersymmetry and without introducing an ad hoc stabilization of DM.

In Fig.\,\ref{fig:mixeddm}, we see that over a wide range of $F_a$ and 
$\lambda_{\chi h}$, one can produce the observed DM abundance easily. 
All of cosmological DM can be axions, if $F_a\approx 10^{12}$\,GeV,  
so that $\Omega_{a}\approx \Omega_{\rm DM}$. A large 
scalar coupling $\lambda_{\chi h}$ suppresses the WIMP density 
$\Omega_{\chi_1^{}}\lesssim10^{-2}\,\Omega_{\rm DM}$. In this limit, 
there is effectively no WIMP DM component and only axion searches are 
expected to be successful.
The other extreme limit is if almost all of DM is comprised of $\chi_1^{}$. 
If $F_a\sim10^9$\,GeV, it suppresses the axion abundance 
to $\Omega_{a}\lesssim10^{-2}\,\Omega_{\rm DM}$ and one can expect 
$\Omega_{\chi}\approx\Omega_{\rm DM}$. This regime is promising for 
traditional WIMP searches, but axion searches wouldn't find a signal. 
An intermediate possibility is to have mixed DM with two components -- axions and 
$\chi_1^{}$. For example, if $F_a\sim\,{\rm few}\times10^{11}\,{\rm GeV}$ and 
$m_1/\lambda_{\chi h}\approx10^3\,$GeV, then $\Omega_{a}\approx\Omega_{\chi_1^{}}\approx
\Omega_{\rm DM}/2$. The phenomenology of this mixed DM can be quite rich.
 We now discuss constraints on and detectability of DM in our scenario.

A strong constraint comes from the invisible width of the 
observed 126 GeV Higgs boson which rules out $\chi_1^{}$ lighter than $m_h/2=$ 62.5\,GeV if 
$\lambda_{\chi h}>10^{-2}$. Bounds from XENON100 also rule out $m_1\lesssim10^{1.9}\,$GeV~\cite{Cline:2013gha}. 
WIMP masses greater than $10\,$TeV require too large values of 
$\lambda_{\chi h}$. We have therefore considered $\chi_1^{}$ in the range 
$100\,{\rm GeV}<m_1<{\rm few}\,$TeV, which restricts the range of 
$\lambda_{\chi h}$ to $\sim(0.1-10)$. $F_a$ is constrained to be 
in the range $(10^9-10^{12})\,$GeV~\cite{Raffelt:2006cw}.

Prospects for detection of DM are very promising. This may be counter-intuitive, because now DM densities of each species are lower and makes it hard to detect them. However, $\chi$ interacts via the Higgs portal at direct detection experiments where there is very high sensitivity. Existing underground experiments, e.g., XENON100 (in 20 yrs), or XENON1T, can probe the entire viable range of $\lambda_{\chi h}$, as long as  WIMPs 
comprise even a few percent of the total DM~\cite{Cline:2013gha}, i.e., for $F_a<$ few$\,\times10^{11}\,$GeV. 
However, indirect detection 
in Fermi, CTA, and Planck 
is possible only if $\chi_1^{}$ forms almost all of DM~\cite{Cline:2013gha} - 
the annihilation signal degrades quadratically for smaller density and evades 
upcoming searches. ADMX is expected to probe 
the axion decay constant $F_a$ in the range $(10^{11}-10^{12})\,$GeV~\cite{Asztalos:2011ei}. 
So, existing direct detection and axion searches will complementarily
 probe  all of the viable parameter space in Fig.\ref{fig:mixeddm}. 
 In other words, a signal in at least one existing experiment is guaranteed.
A smoking gun signature of mixed DM would be signals for both direct 
detection searches and axion searches.

{\bf{\emph{Neutrino Mass in Axion Models.---}}} The KSVZ model has 
heavy quarks $Q_{L,R}$ and a complex scalar $\zeta$. We 
added the scalar $\chi$ as the dark matter candidate. Neutrino mass may be generated 
radiatively in these models, if the new particles charged under $U(1)_{PQ}$ are added. 
We provide two concrete realizations of this idea.

{{\emph{Model I.---}}}
To get neutrino masses, we only add a neutral singlet fermion $N_R$ (per generation) 
and a new scalar doublet $\eta = (\eta^+,\eta^0)^T$ with $\eta^0 = (\eta_1^{}+i\, \eta_2^{})/\sqrt2$, 
all of which transform under $U(1)_{PQ}$. Quantum numbers of the new particles 
are listed in Table~\ref{Tab:1-loop-PQ}.

\begin{table}[!h]
\caption{New particles in the one-loop radiative seesaw model with Peccei-Quinn 
symmetry.}
\begin{ruledtabular}
\label{Tab:1-loop-PQ}
\begin{spacing}{1.2}
\begin{tabular}{l c c c c c c}
 {}			& $Q_L$ 				& $Q_R$ 				& $\zeta$ 	& $\chi$ 	& $N_R$ 				& $\eta$ 			\\ \hline
 spin 		& $\nicefrac{1}{2}$ 	& $\nicefrac{1}{2}$ 	& $0$ 		& $0$		& $\nicefrac{1}{2}$ 	& $0$ 				\\
 $SU(3)_c$ 	& ${\bf 3}$			& ${\bf 3}$			& ${\bf 1}$	& ${\bf 1}$ 	& ${\bf 1}$ 			& ${\bf 1}$ 			\\
 $SU(2)_L$ 	& ${\bf 1}$ 			& ${\bf 1}$ 			& ${\bf 1}$ 	& ${\bf 1}$ 	& ${\bf 1}$ 			& ${\bf 2}$ 			\\
 $U(1)_Y$ 	& $\nicefrac{-1}{3}$ & $\nicefrac{-1}{3}$ & $0$ 		& $0$ 		& $0$ 				& $\nicefrac{1}{2}$ 	\\
 $U(1)_{PQ}$	& $1$ 				& $-1$ 				& $2$ 		& $1$		& $1$ 				& $1$ 				
\end{tabular}
\end{spacing}
\end{ruledtabular}
\end{table}

\begin{figure}[!t]
\centering
\includegraphics[width=6cm]{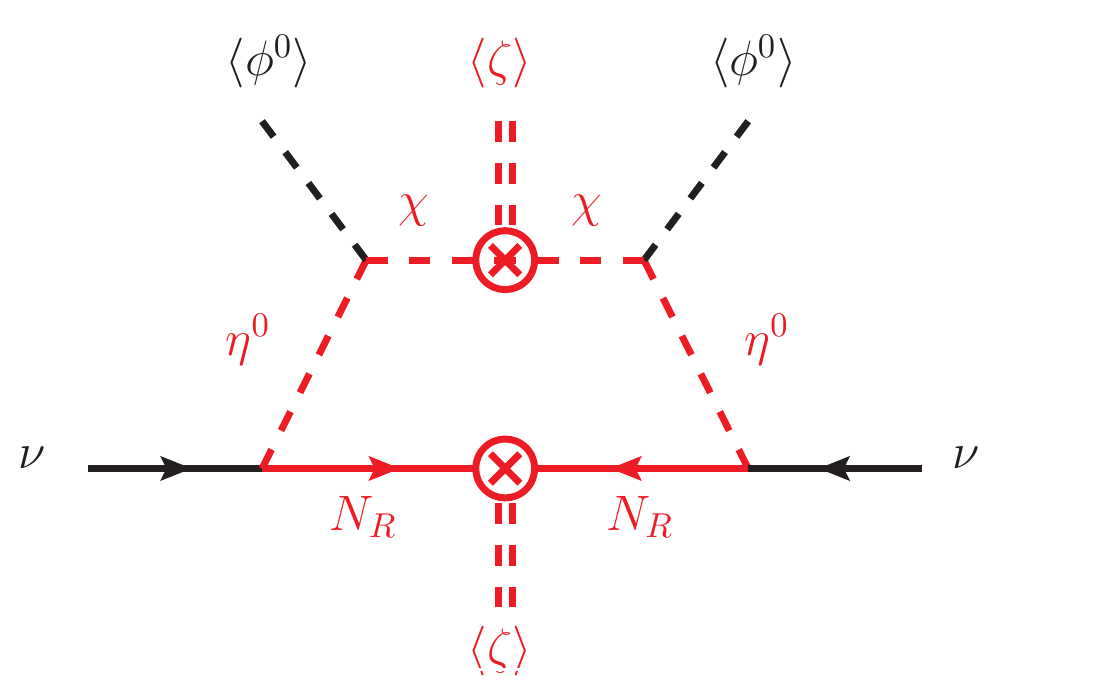}
\caption{Diagram for the one-loop radiative neutrino mass. }
\label{FIG:1-loop-PQ}
\end{figure}

Radiative neutrino mass is then generated in one loop as shown in Fig.\,\ref{FIG:1-loop-PQ}, in 
analogy to the original $Z_2$ scotogenic model~\cite{Ma:2006km} as $\zeta$ 
acquires a VEV, thus breaking $U(1)_{PQ}$ to $Z_2$. 

The particles $Q,\,\chi_{1,2},\,\eta_{1,2},\,\eta^\pm,\,{\rm and}\,N_i$ are odd under $Z_2$, 
whereas all others (including $\sigma$ and $a$) are even. 
Although $\sigma$ mixes with $h$, they are almost mass eigenstates because 
$v_{\rm SM} \ll F_a$.   As for $\chi_{1,2}$ and $\eta_{1,2}$, they are completely 
mixed in a $4 \times 4$ matrix (including the $\Phi^\dagger \eta \chi \zeta^*$ 
term not shown in Fig.\,\ref{FIG:1-loop-PQ}), the lightest of which is now 
the WIMP-DM candidate.  

However, the radiative neutrino 
mass is still of the generic form
\begin{equation}
({\cal M})_{ij} = \sum_{k} \frac{h_{ik} h_{jk} M_k}{16 \pi^2}
\sum_\alpha \frac{(U_{1 \alpha}^2 - U_{2 \alpha}^2)m^2_\alpha}{m^2_\alpha - M_k^2} 
\ln{\frac{m_\alpha^2}{M_k^2}},
\end{equation}
where $U_{1 \alpha}$ and $U_{2 \alpha}$ are the unitary matrices which link 
$\eta_{1,2}$ to the four mass eigenstates of mass $m_\alpha$, $h_{ij}$ are the 
Yukawa couplings, and $M_k$ are the heavy neutrino masses. Note that 
in the original model~\cite{Ma:2006km}, there are only two mass eigenstates 
with $U_{11} = U_{22} =1$ and $U_{12} = U_{21} = 0$. 
Radiative lepton flavor violation (LFV) $\ell_i \to \ell_j \gamma$ is induced 
in general by $\eta^\pm$ exchange, which may be 
suppressed by small $h_{ik}$, as in Ref.\,~\cite{Ma:2006km}.

{{\emph{Model II.---}}}
Another interesting possibility is to consider scalar leptoquarks and diquarks 
transforming under $U(1)_{PQ}$. Quantum numbers of the new particles 
are listed in Table~\ref{Tab:2-loop-PQ}.

\begin{figure}[!t]
\centering
\includegraphics[width=6cm]{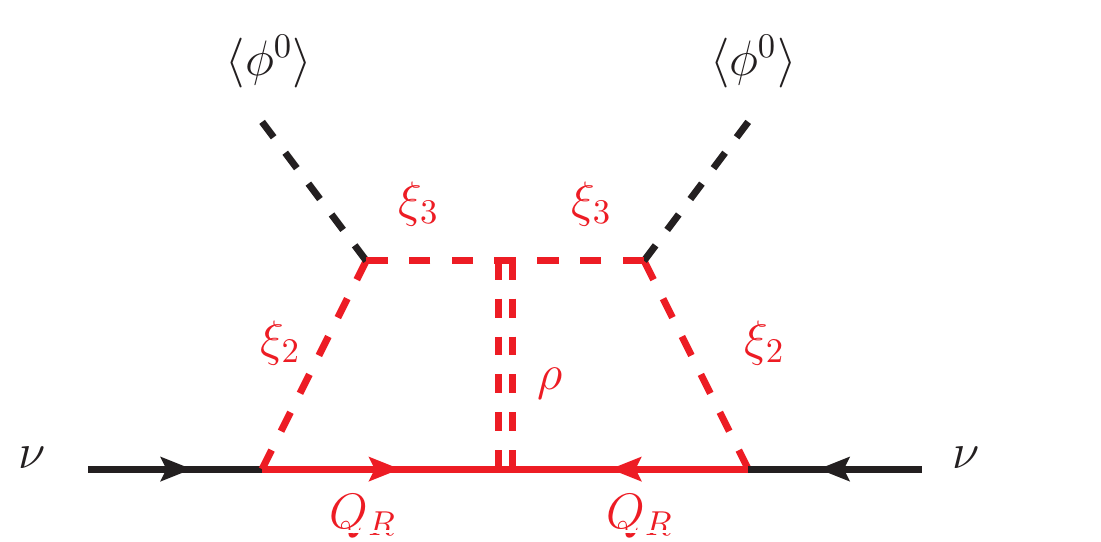}
\caption{Diagram for the two-loop radiative neutrino mass. }
\label{FIG:2-loop-PQ}
\end{figure}

\begin{table}[!h]
\begin{ruledtabular}
\caption{New particles in the two-loop radiative seesaw model with the Peccei-Quinn 
symmetry.}
\label{Tab:2-loop-PQ}
\begin{spacing}{1.2}
\begin{tabular}{lccccccc}
 {} 			& $Q_L$ 				& $Q_R$ 				& $\zeta$ 	& $\chi$ 	& $(\xi_1,\xi_2)$ 	& $\xi_3$ 			& $\rho$\\\hline
spin 		& $\nicefrac{1}{2}$ 	& \nicefrac{1}{2} 	& $0$ 		& $0$		& $0$ 				& $0$				& $0$ 			\\
 $SU(3)_c$	& ${\bf 3}$			& ${\bf 3}$ 			& ${\bf 1}$ 	& ${\bf 1}$ 	& ${\bf 3}$ 			& ${\bf 3}$ 			& ${\bf 6}$ 			\\
 $SU(2)_L$ 	& ${\bf 1}$ 			& ${\bf 1}$ 			& ${\bf 1}$ 	& ${\bf 1}$ 	& ${\bf 2}$ 			& ${\bf 1}$			& ${\bf 1}$ 			\\
 $U(1)_Y$	& $\nicefrac{-1}{3}$ & \nicefrac{-1}{3} 	& $0$ 		& $0$ 		& $\nicefrac{1}{6}$ 	& \nicefrac{-1}{3} 	& \nicefrac{-2}{3} 	\\
 $U(1)_{PQ}$	& $1$ 				& $-1$ 				& $2$ 		& $1$		& $-1$ 				& $-1$ 				& $-2$ 				
 \end{tabular}
\end{spacing}
\end{ruledtabular}
\end{table}
%
Radiative neutrino mass is then generated in two loops as shown in Fig.\,\ref{FIG:2-loop-PQ}, in 
analogy with the recent proposal of Ref.~\cite{Kohda:2012sr}. Note the remarkable result 
that a Majorana neutrino mass is radiatively generated without breaking 
$U(1)_{PQ}$. 
%
%
The Lagrangian relevant for the extended sector is given by
\begin{eqnarray}
{\cal L} &=& y_Q^{} \bar{Q}_R \nu_L \xi_2 
+ h_{QQ'}^{} \rho^* Q_R Q'_R \nonumber \\  
&-& \epsilon_\xi^{} \phi^0 \xi_2^* \xi_3 - \epsilon_\rho^{} \rho^* \xi_3 \xi_3 + \text{H.c}.
\end{eqnarray}
The $\epsilon_\xi^{}$ term mixes $\xi_2$ and $\xi_3$ with angle $\theta_\xi^{}$ to 
form mass eigenstates.
The two-loop neutrino mass matrix is then calculated as
\begin{align}
&({\cal M})_{ij} = \sum_{Q, Q'} 8 h_{QQ'}^{}
\sum_{\alpha, \beta} \kappa_{\alpha\beta}\,  
y_Q^i\, m_\alpha\,  I_{\alpha \beta}^{QQ'}\, m_\beta\, y_{Q'}^j, 
%
\end{align}
where
\begin{align}
\kappa_{\alpha\beta}
&= \epsilon_\rho^{}
\begin{pmatrix} \sin^2\theta_\xi^{} & \sin\theta_\xi^{} \cos\theta_\xi^{} \\ 
\sin\theta_\xi^{} \cos\theta_\xi^{} & \cos^2\theta_\xi^{} \end{pmatrix},
\end{align}
%
\begin{align}
I_{\alpha\beta}^{QQ'}
&= 
+\int \frac{d^4k_1}{(2\pi)^4} 
\int \frac{d^4k_2}{(2\pi)^4} 
\frac1{k_1^2-M_Q^2} 
\frac1{k_2^2-M_{Q'}^2} 
\nonumber \\
&
\quad \times
\frac1{(k_1+k_2)^2-M_\rho^2} 
\frac1{k_1^2-m_\alpha^2} 
\frac1{k_2^2-m_\beta^2}.
\end{align}
The LFV process, $\ell_i \to \ell_j \gamma$, is induced by the  
$\xi_1^{2/3}$ leptoquark.  
These branching fractions could be easily suppressed by choosing relatively 
small Yukawa coupling $y_Q^{}$ without making the two-loop neutrino 
mass too small.  This would have been difficult if a three-loop neutrino mass were  
considered.

{\bf{\emph{Collider Phenomenology.---}}}
While the scale of $U(1)_{PQ}$ symmetry breaking must be very high, the 
KSVZ singlet quark $Q$ may be light enough to be copiously produced at the 
Large Hadron Collider (LHC) via $gg \to Q \bar Q$.   Once produced, it decays 
into a $d$ quark and 
either $\chi_1^{}$ or $\chi_2$.  Whereas $\chi_1^{}$ appears as missing energy, 
$\chi_2$ decays to $\chi_1^{} d \bar{d}$.  Similar studies where a heavy quark 
decays into a top quark plus DM have appeared~\cite{Alwall:2010jc}, 
and its experimental search at the LHC reported~\cite{Aad:2011wc}. 
Although we have assumed specifically that $Q$ has charge $-1/3$, our model 
is easily adapted to $2/3$ as well.  Future LHC analysis of such heavy 
quark decays will be important in testing our proposal.
For example, the exclusive search for supersymmetric scalar quarks 
may be reinterpreted as mass bounds on $Q$. 

In Model II, we have additional signals at colliders. There can be copious 
production of the leptoquarks and diquarks also via 
$gg \to \xi^{+1/3}\xi^{-1/3}, \xi^{+2/3}\xi^{-2/3}, \rho^{-2/3}\rho^{+2/3}$.  
There are many possible decay chains.  For example, 
$\xi^{2/3}$ may decay into a charged lepton plus $Q^{-1/3}$ with the latter 
decaying into $d$ and $\chi_1^{}$.  This may contaminate $t \bar{t}$ pair 
production with $t \to b W^+ \to b\ell^+ \nu$.  
The reinterpretation of $t\bar{t}$ events may give a constraint 
on $\xi^{2/3}$. This phenomenology is rich, and we leave it for further study.
%

{\bf{\emph{Conclusion.---}}}
We have proposed a unified framework for solving three outstanding problems in 
particle physics and astrophysics.  We invoke the usual Peccei-Quinn symmetry 
to solve the strong CP problem, resulting in a very light axion.  However, 
we also make the simple (but hitherto unexplored) observation that in all 
axion models, $U(1)_{PQ}$ also leaves a residual $Z_2$ symmetry, and in the 
KSVZ model, it may be used for stabilizing dark matter. In other words, 
DM is stability is related to the absence of strong CP violation.  We make the minimal addition of a complex scalar field $\chi = (\chi_1^{} + i \chi_2)/\sqrt{2}$ to the the KSVZ model with 
the interaction $\chi \bar{Q}_L d_R^{}$ as well as the usual extra terms 
which appear in the Higgs potential.  Consequently, $\chi_1$ behaves 
naturally as the singlet scalar in the well-studied simplest of all possible 
dark-matter models.  In other words, we have provided a theoretical 
justification for this otherwise {\it ad hoc} proposal. Phenomenologically,  
our scenario is extremely promising, with guaranteed signals at 
direct-detection experiments or axion searches (or both). The same $Z_2$ symmetry 
may also be connected to the well-studied notion of radiative neutrino mass 
through dark matter. To implement this notion of radiative neutrino mass, new particles are 
required, which are charged under $U(1)_{PQ}$.  Collider searches for 
these new particles are also promising, especially Model II where 
leptoquark and diquark scalars may be produced copiously at the LHC.

{\bf{\emph{Acknowldgements.---}}}
E.\,M. thanks 
the International Centre for Theoretical Physics for their hospitality. 
His work is supported in part by the U.~S.~Department of Energy under 
Grant No.~DE-FG03-94ER40837.
 K.T.\ was supported, in part, by the Grant-in-Aid for Scientific research
from the Ministry of Education, Science, Sports, and Culture~(MEXT), Japan,
No.~23104011. 

\vspace{0cm}
\bibliographystyle{apsrev}
\interlinepenalty=10000
\tolerance=100
\bibliography{Bibliography/references}

\end{document}